\documentstyle[psfig]{l-aa}
\def \etal     {et al.}
\def \eg       {e.\,g.}
\def \vLSR     {\hbox{${v_{\rm LSR}}$}}
\def \delv     {\hbox{$\Delta v_{1/2}$}}
\def \TMB      {\hbox{$T_{\rm MB}$}}
\def \Tkin     {\hbox{$T_{\rm kin}$}}
\def \Tsys     {\hbox{$T_{\rm sys}$}}
\def \Tdust    {\hbox{$T_{\rm dust}$}}
\def \Mgas     {\hbox{$M_{\rm gas}$}}                     
\def \Msol     {\hbox{M$_{\odot}$}}                       
\def \Mvir     {\hbox{$M_{\rm vir}$}}                     
\def \numd     {\hbox{$n\,({\rm H}_2$)}}
\def \solum    {\hbox{$L_{\odot}$}}
\def \HII      {H\,{\sc ii}}
\def \kms      {\hbox{${\rm km\,s}^{-1}$}}                
\def \Kkms     {\hbox{${\rm K\,km\,s}^{-1}$}}             
\def \percc    {\hbox{${\rm cm}^{-3}$}}                   
\def \cmsq     {\hbox{${\rm cm}^{-2}$}}                   
\def \arcdeg   {\hbox{$^{\circ}$}}                        
\def \arcmin   {\hbox{$^\prime$}}                         
\def \arcsec   {\hbox{$^{\prime\prime}$}}                 
\def \MOLH     {\hbox{H$_2$}}                             
\def \CCH      {\hbox{C$_2$H}}                            
\def \twCO     {\hbox{$^{12}$CO}}                         
\def \thCO     {\hbox{$^{13}$CO}}                         
\def \CseO     {\hbox{C$^{17}$O}}                         
\def \CeiO     {\hbox{C$^{18}$O}}                         
\def \CtwS     {\hbox{C$^{32}$S}}                         
\def \CfoS     {\hbox{C$^{34}$S}}                         
\def \FORM     {\hbox{H$_2$CO}}                           
\def \METH     {\hbox{CH$_3$OH}}                          
\def \CYCP     {\hbox{C$_3$H$_2$}}                        
\def \HCOp     {\hbox{HCO$^+$}}                           
\def \C#1      {\hbox{$^{#1}$C}}                          
\def \N#1      {\hbox{$^{#1}$N}}                          
\def \O#1      {\hbox{$^{#1}$O}}                          
\def \S#1      {\hbox{$^{#1}$S}}                          
\newcommand{\see}[1]{$^{\rm #1)}$}
\newcommand{\vol}[1]{{\rm #1}}
\def \ARAA     {{\rm ARA\&A}}
\def \AaA      {{\rm A\&A}}
\def \AaAS     {{\rm A\&AS}}
\def \AaAR     {{\rm A\&AR}}
\def \ApJ      {{\rm ApJ}}
\def \ApJS     {{\rm ApJS}}
\def \MNRAS    {{\rm MNRAS}}
\def \uspace#1 {\makebox{\rule[#1ex]{0pt}{2ex}}}
\def \dspace   {\makebox{\rule[-2ex]{0pt}{2ex}}}

\begin{document}

\thesaurus{11 (09.01.1; 09.13.2; 11.01.1; 11.09.4; 11.13.1; 13.19.3) }

\title{Molecular abundances in the Magellanic Clouds}
\subtitle{III. LIRS\,36, a star-forming region in the SMC
          \thanks {Based on observations with the Swedish-ESO
          Submillimeter Telescope (SEST) at the European Southern
          Observatory (ESO), La Silla, Chile } }

\author{Y.-N. Chin\inst{1,2}, C.~Henkel\inst{3,4}, T.J.~Millar\inst{5},
        J.B.~Whiteoak\inst{6,7}, \and M.~Marx-Zimmer\inst{2}}

\offprints{Y.-N. Chin, ASIAA, Taiwan, einmann@biaa21.biaa.sinica.edu.tw}

\institute{
   Institut of Astronomy and Astrophysics, Academia Sinica,
   P.O.Box 1-87, Nankang, Taipei, Taiwan
\and
   Radioastronomisches Institut der Universit\"at Bonn,
   Auf dem H\"ugel 71, D-53121 Bonn, Germany
\and
   Max-Planck-Institut f\"ur Radioastronomie,
   Auf dem H\"ugel 69, D-53121 Bonn, Germany
\and
   European Southern Observatory, Casilla 19001,
   Santiago 19, Chile
\and
   Department of Physics, UMIST,
   P O Box 88, Manchester M60 1QD, United Kingdom
\and
   Australia Telescope National Facility, Radiophysics Laboratories,
   P.O. Box 76, Epping, NSW 2121, Australia
\and
   Paul Wild Observatory, Australia Telescope National Facility, CSIRO,
   Locked Bag 194, Narrabri NSW 2390, Australia
}

\date{Received date / Accepted date}

\maketitle

\begin{abstract}

   Detections of CO, CS, SO, \CCH, \HCOp, HCN, HNC, \FORM,
   and \CYCP\ are reported from LIRS\,36,
   a star-forming region in the Small Magellanic Cloud.
   \CeiO, NO, \METH, and most notably CN have not been detected,
   while the rare isotopes \thCO\ and, tentatively, \CfoS\ are seen.
   This is so far the most extensive molecular multiline study
   of an interstellar medium with a heavy element depletion
   exceeding a factor of four.

   The $X$ = $N$(H$_2$)/$I_{\rm CO}$ conversion factor is
   $\approx$ $4.8 \times 10^{21}$\,\cmsq\,(\Kkms)$^{-1}$, slightly
   larger than the local Galactic disk value.
   The CO (1--0) beam averaged column density then becomes $N$(\MOLH) $\approx$
   $3.7 \times 10^{21}$\,\cmsq\ and the density \numd\ $\approx$ 100\,\percc.
   A comparison with $X$-values from Rubio \etal\ (1993a) shows that
   on small scales ($R$ $\approx$ 10\,pc) $X$-values are more similar to
   Galactic disk values than previously anticipated, favoring a neutral
   interstellar medium of predominantly molecular nature in the cores.
   The $I$(\thCO)/$I$(\CeiO) line intensity ratio
   indicates an underabundance of $^{12}$C$^{18}$O relative to
   $^{13}$C$^{16}$O w.r.t.\ Galactic clouds.
   $I$(\HCOp)/$I$(HCN) and $I$(HCN)/$I$(HNC) line intensity ratios are
   $>$ 1 and trace a warm (\Tkin\ $>$ 10\,K) molecular gas
   exposed to a high ionizing flux.
   Detections of the CS $J$=2--1, 3--2, and 5--4 lines imply the presence of
   a high density core with \numd\ = $10^5 - 10^7$\,\percc.
   In contrast to star-forming regions in the LMC,
   the CN 1--0 line is substantially weaker than the corresponding
   ground rotational transitions of HCN, HNC, and CS.
   CO, CS, \HCOp, and \FORM\ fractional abundances are a factor $\approx$ 10
   smaller than corresponding values in Galactic disk clouds.
   Fractional abundances of HCN, HNC, and likely CN are even
   two orders of magnitude below their `normal', Galactic disk values.
   The CN/CS abundance ratio is $\la$ 1.
   Based on chemical model calculations,
   we suggest that this is because of the small metallicity of the SMC,
   which affects the destruction of CN but not CS, and because of
   the high molecular core density which also favors CN destruction.


\end{abstract}

\section{Introduction}
 \label{sec:MC3-Introduction}

   Recent studies of the dense molecular gas in five star-forming regions of the
   Large Magellanic Cloud (LMC) revealed a number of striking differences with
   respect to properties typically observed in the clouds of the Galactic disk
   (see Johansson \etal\ 1994; Chin \etal\ 1997):
   \CeiO\ is underabundant relative to \thCO, the \CeiO/\CseO\ ratio
   ($\approx$ 2) appears to be smaller than the canonical value of 3.5
   in the Galactic interstellar medium, and \HCOp\ to HCN line
   intensity ratios are larger than those in the Milky Way.
   In view of these results, one may expect that a thorough study of
   the Small Magellanic Cloud (SMC) with its low metallicity and strong UV
   radiation field will reveal even more drastic deviations from typical
   Galactic disk properties, thus permitting insights into otherwise not
   directly discernible astrophysical and astrochemical processes
   (cf.\ Johansson 1991, 1997; Rubio \etal\ 1993a,b, 1997; Lequeux \etal\ 1994;
   Chin \etal\ 1997; Heikkil\"a \etal\ 1997).
   Employing a Schottky receiver, Chin \etal\ (1997) reported the
   detection of \CCH, CS, and a tentative detection of CN toward LIRS\,36,
   the source with strongest CO $J$=1--0 line temperature among investigated
   SMC IRAS sources (Rubio \etal\ 1993b).

   Here, a high sensitivity study of LIRS\,36 is presented, providing a detailed
   molecular view of a star formation region in the extremely metal poor
   environment (\eg\ Westerlund 1990) that characterizes the SMC.

\begin{figure}
   \vspace*{-25 mm}
   \hspace*{-2 mm} \psfig{figure=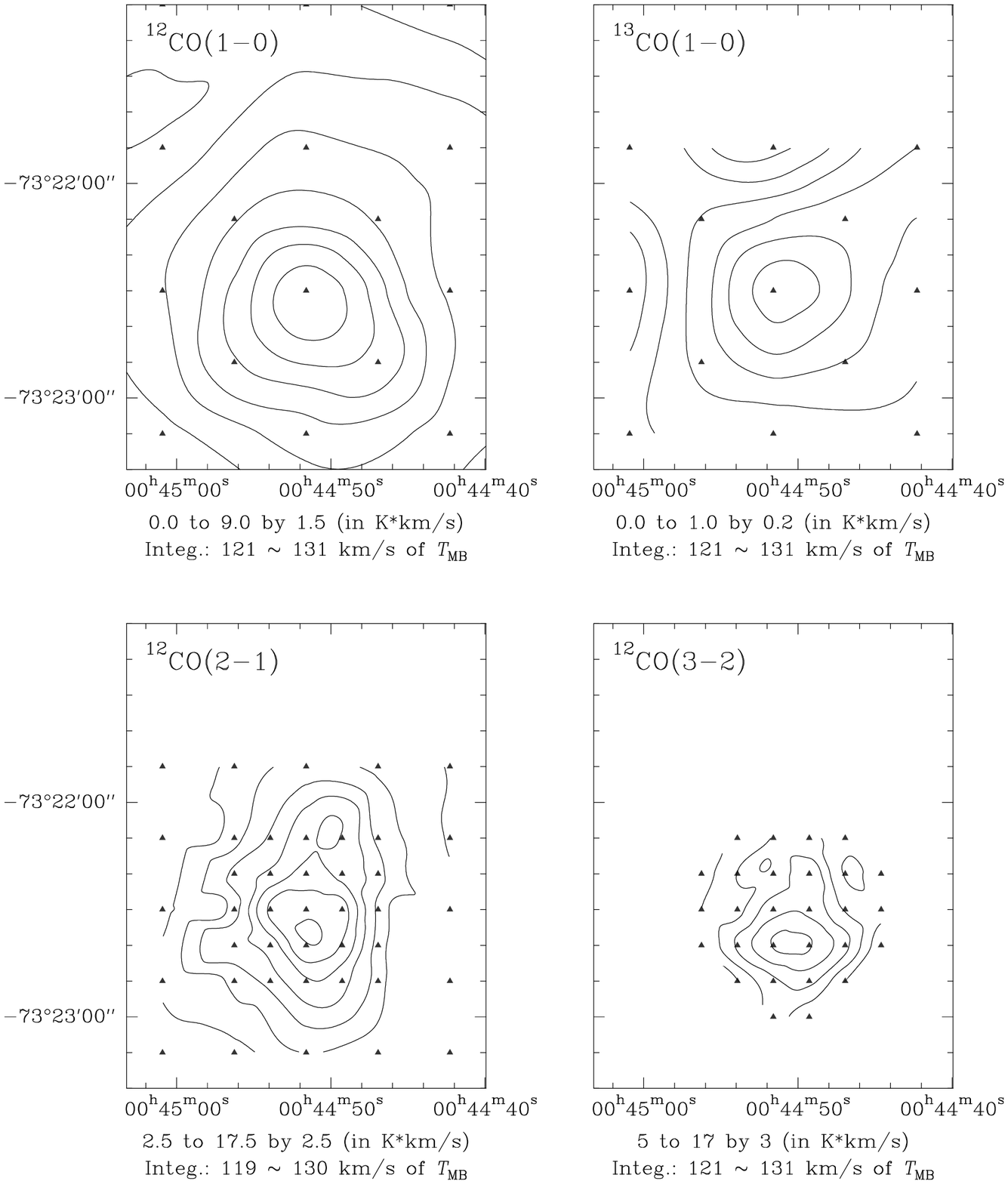,width=95mm,angle=0}
   \vspace*{-10 mm}
   \caption[]
           {Contour maps of LIRS36.
            {\bf a)} \twCO(1--0): contour levels are
             0.0 to 9.0 by 1.5\,\Kkms\ integrated
             between \vLSR\ = 121 and 131 \kms.
            {\bf b)} \thCO(1--0): contour levels are
             0.0 to 1.0 by 0.2\,\Kkms\ integrated
             between \vLSR\ = 121 and 131 \kms.
            {\bf c)} \twCO(2--1): contour levels are
             2.5 to 17.5 by 2.5\,\Kkms\ integrated between
             \vLSR\ = 119 and 130 \kms.
            {\bf d)} \twCO(3--2): contour levels are
             5 to 17 by 3\,\Kkms\ integrated between
             \vLSR\ = 121 and 131 \kms.
            Typical r.m.s.\ values are 0.14, 0.25, and 0.3\,\Kkms, respectively.
           }
 \label{fig:contour-LIRS36}
\end{figure}

\begin{figure*}
   \vspace*{-30 mm}
   \hspace*{5 mm} \psfig{figure=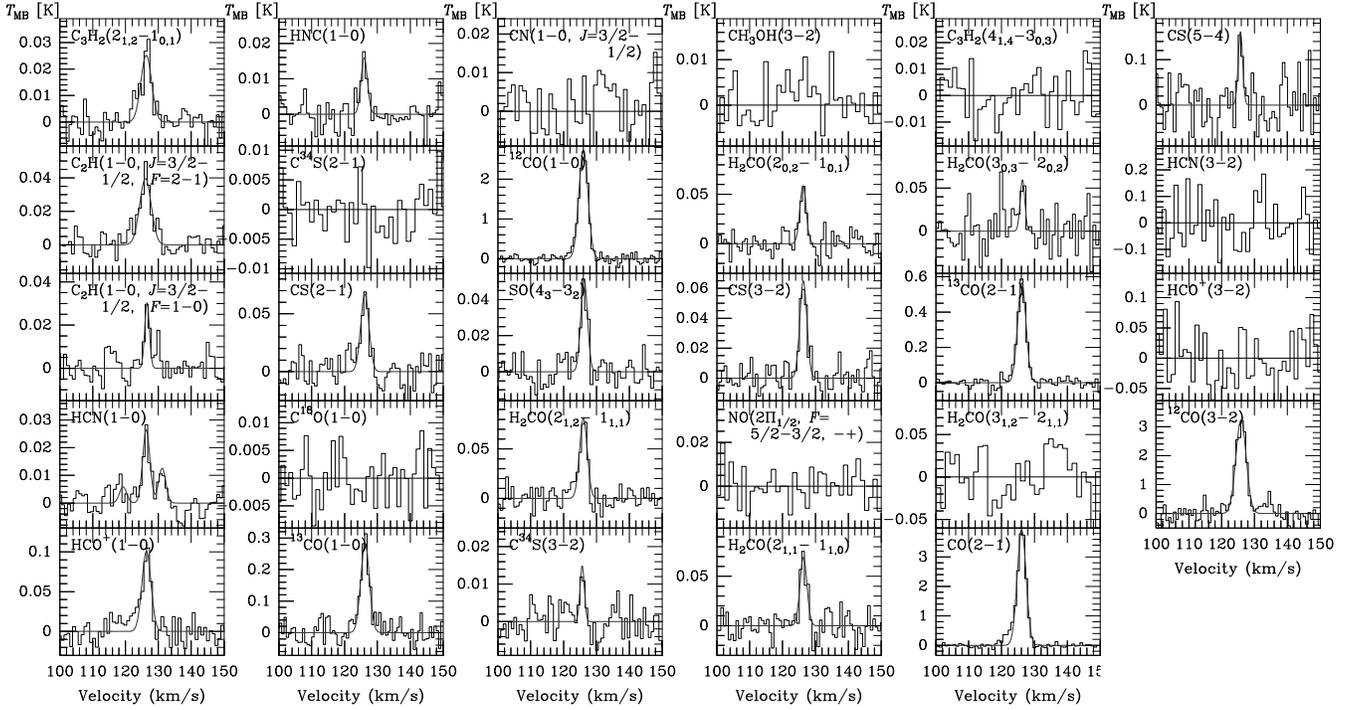,width=19 cm,angle=90}
   \vspace*{-25 mm}
   \caption[]
           {Molecular spectra toward LIRS\,36, with the line rest frequency
            increasing from the upper left to the lower right spectrum.
            Observed nominal peak position:
            $\alpha_{1950}$ = 00$^{\rm h}$44$^{\rm m}$51$^{\rm s}$,
            $\delta_{1950}$ = $-$73\arcdeg 22\arcmin 30\arcsec.}
 \label{fig:LIRS36}
\end{figure*}

\section{Observations}
 \label{sec:MC3-Observations}

   The data were taken in January 1996 and 1997 and in July 1997,
   using the 15-m Swedish-ESO Submillimetre Telescope (SEST) at La Silla, Chile.
   Two SIS receivers, one at $\lambda$ = 3\,mm and one at 2\,mm,
   yielded overall system temperatures, including sky noise,
   of order \Tsys\ = 250\,K on a main beam brightness temperature (\TMB)
   scale while for \twCO(1-0) \Tsys\ reached 400\,K.
   In January 1997, a $\lambda$ = 1.3\,mm SIS receiver yielded overall
   \TMB\ system temperatures of 500 -- 1600\,K at 218 -- 245 GHz and
   3000 -- 4000\,K at 265 -- 267\,GHz in clear but humid weather.
   In July 1997, a 0.85\,mm SIS receiver
   with \Tsys\ $\approx$ 3000\,K was employed.
   The backend was an acousto-optical spectrometer (AOS)
   which was split into 2 $\times$ 1000 contiguous channels for simultaneous
   $\lambda$ $\approx$ 3 and 2\,mm observations.
   At $\lambda$ $\approx$ 1.3 and 0.85\,mm, all 2000 channels were used to
   cover a similar velocity range.
   The channel separation of 43\,kHz corresponds to 0.04, 0.05 -- 0.06,
   0.08 -- 0.10 and 0.11 -- 0.15\,\kms\ for the frequency intervals 340 -- 350,
   265 -- 220, 150 -- 130 and 115 -- 85\,GHz, respectively.
   Depending on line frequency, the antenna beamwidth varied
   from 15\arcsec\ to 57\arcsec\ corresponding to linear scales of 4 -- 17\,pc.
   At 345\,GHz, the beam was slightly broadened in E--W direction;
   this is caused by astigmatism, reflecting the limited surface
   accuracy of the SEST antenna (L.-\AA.~Nyman, priv.~comm.).

   The observations were carried out in a dual beam-switching mode (switching
   frequency 6\,Hz) with a beam throw of 11\arcmin 40\arcsec\ in azimuth.
   The on-source integration time of each spectrum varied from 8 minutes
   for \twCO(1-0) to 18.5 hours for \CfoS(3-2).
   All spectral intensities were converted to a \TMB\ scale, correcting for
   main beam efficiencies of 0.74 at 85--100, 0.70 at 100--115,
   0.67 at 130--150, 0.45 at 220--265\,GHz, and 0.30 at 345\,GHz
   (L.B.G.~Knee, L.-\AA.~Nyman, A.R.~Tieftrunk, priv.~comm.).
   Calibration was checked by measurements of Orion KL, M17SW, and NGC\,4945;
   it was found to be accurate to $\pm$ 10\%.
   The pointing accuracy, obtained from measurements of the SiO masers
   R\,Dor and U\,Men, was better than 10\arcsec.
   345\,GHz data with pointing offsets in excess of 5\arcsec\
   were ignored in the data analysis.

\begin{table*}
   \caption[]
           {Parameters of the Observed Molecular Lines toward LIRS\,36}
 \label{tbl:LIRS36}
 \scriptsize
   \begin{flushleft}
   \begin{tabular}{l l r c r c c r@{.}l@{~$\pm$~}l}
   \hline
   \multicolumn{2}{l}{Molecule}      & Frequency & \TMB \see{a} \uspace1
        & \multicolumn{1}{c}{r.m.s.\see{b} \hspace*{-5mm}} & \vLSR  & \delv
        & \multicolumn{3}{c}{$\int$ \TMB\,d$v$ \see{c}} \\
   \multicolumn{2}{l}{\& Transition} & [MHz]~~~  & [K]   \dspace
        & \multicolumn{1}{c}{[mK]   \hspace*{-4mm}} & [\kms] & [\kms]
        & \multicolumn{3}{c}{[\Kkms]}           \\
   \hline
   \uspace1
   \CYCP          & $J$=2$_{1,2}$--1$_{0,1}$        &  85338.890
                  & 0.025 &   7 &  126.3 & 3.9 &  0&108 & 0.008 \\
   \uspace0.5
   \CCH           & $N$=1-0 $J$=3/2--1/2 $F$=2--1   &  87316.925
                  & 0.040 &   8 &  126.0 & 4.1 &  0&180 & 0.008 \\
   \uspace0.5
                  & \hspace*{23.2mm}     $F$=1--0   &  87328.624
                  & 0.030 &   7 &  126.6 & 1.8 &  0&075 & 0.008 \\
   \uspace0.5
   HCN            & $J$=1--0                        &  88631.847
                  & 0.026 &   5 &  126.3 & 2.5 &  0&106 & 0.006 \\
   \uspace0.5
   \HCOp          & $J$=1--0                        &  89188.518
                  & 0.101 &  17 &  126.4 & 3.1 &  0&369 & 0.018 \\
   \uspace0.5
   HNC            & $J$=1--0                        &  90663.543
                  & 0.016 &   5 &  126.0 & 2.4 &  0&038 & 0.005 \\
   \uspace0.5
   \CfoS          & $J$=2--1                        &  96412.982
              & $<$ 0.01  &   7 & \ldots & \ldots
                        & \multicolumn{3}{c}{$<$ 0.026}         \\
   \uspace0.5
   CS             & $J$=2--1                        &  97980.968
                  & 0.067 &  16 &  126.2 & 2.8 &  0&225 & 0.016 \\
   \uspace0.5
   \CeiO          & $J$=1--0                        & 109782.160
              & $<$ 0.01  &   8 & \ldots & \ldots
                        & \multicolumn{3}{c}{$<$ 0.030}         \\
   \uspace0.5
   \thCO          & $J$=1--0                        & 110201.353
                  & 0.284 &  43 &  126.2 & 3.1 &  1&04  & 0.05  \\
   \uspace0.5
   CN             & $N$=1--0 $J$=3/2--1/2 $F$=5/2--3/2
                                                    & 113490.982
              & $<$ 0.01  &  12 & \ldots & \ldots
                        & \multicolumn{3}{c}{$<$ 0.038}         \\
   \uspace0.5
   CO             & $J$=1--0                        & 115271.204
                  & 2.73  & 135 &  126.1 & 3.2 &  9&69  & 0.14  \\
   \uspace0.5
   SO             & $J$=4$_3$--3$_2$                & 138178.648
                  & 0.051 &  11 &  126.3 & 2.3 &  0&135 & 0.010 \\
   \uspace0.5
   \FORM          & $J$=2$_{1,2}$--1$_{1,1}$        & 140839.518
                  & 0.081 &  16 &  126.1 & 2.7 &  0&241 & 0.014 \\
   \uspace0.5
   \CfoS          & $J$=3--2 \see{d}                & 144617.147
                  & 0.015 &   7 &  125.3 & 1.9 &  0&018 & 0.007 \\
   \uspace0.5
   \METH          & $J$=3$_0$--2$_0$, A$^+$         & 145103.230
              & $<$ 0.01  &  10 & \ldots & \ldots
                        & \multicolumn{3}{c}{$<$ 0.029}         \\
   \uspace0.5
   \FORM          & $J$=2$_{0,2}$--1$_{0,1}$        & 145602.953
                  & 0.059 &  16 &  126.2 & 2.2 &  0&142 & 0.013 \\
   \uspace0.5
   CS             & $J$=3--2                        & 146969.049
                  & 0.065 &  14 &  126.3 & 2.5 &  0&179 & 0.012 \\
   \uspace0.5
   NO             & $2\Pi_{1/2}$ $J$=3/2--1/2 $F$=5/2--3/2 ($-$+)
                                                    & 150176.480
              & $<$ 0.01  &  17 &  \ldots & \ldots
                        & \multicolumn{3}{c}{$<$ 0.048}         \\
   \uspace0.5
   \FORM          & $J$=2$_{1,1}$--1$_{1,0}$        & 150498.339
                  & 0.069 &  19 &  126.4 & 2.5 &  0&167 & 0.016 \\
   \uspace0.5
   \CYCP          & $J$=4$_{1,4}$--3$_{0,3}$        & 150851.910
              & $<$ 0.02  &  20 & \ldots & \ldots
                        & \multicolumn{3}{c}{$<$ 0.055}         \\
   \uspace0.5
   \FORM          & $J$=3$_{0,3}$--2$_{0,2}$ \see{d}& 218222.191
                  & 0.060 &  44 &  126.4 & 1.7 &  0&150 & 0.034 \\
   \uspace0.5
   \thCO          & $J$=2--1                        & 220398.686
                  & 0.565 &  48 &  126.1 & 3.0 &  1&85  & 0.03  \\
   \uspace0.5
   \FORM          & $J$=3$_{1,2}$--2$_{1,1}$        & 225697.772
              & $<$ 0.04  &  74 & \ldots & \ldots
                         & \multicolumn{3}{c}{$<$ 0.168}         \\
   \uspace0.5
   CO             & $J$=2--1                        & 230537.990
                  & 3.92  & 121 &  126.1 & 3.2 & 13&9   & 0.2   \\
   \uspace0.5
                  & $J$=2--1 \see{e}                & 230537.990
                  & 3.02  & 112 &  126.1 & 3.5 & 11&7   & 0.1   \\
   \uspace0.5
   CS             & $J$=5--4                        & 244935.606
                  & 0.161 &  93 &  125.7 & 1.6 &  0&287 & 0.067 \\
   \uspace0.5
   HCN            & $J$=3--2                        & 265886.432
              & $<$ 0.1   & 250 & \ldots & \ldots
                        & \multicolumn{3}{c}{$<$ 0.52}          \\
   \uspace0.5
   \HCOp          & $J$=3--2                        & 267557.625
              & $<$ 0.1   & 120 & \ldots & \ldots
                        & \multicolumn{3}{c}{$<$ 0.25}          \\
   \uspace0.5
   CO             & $J$=3--2                        & 345795.975
                  & 3.23  & 570 &  125.9 & 3.7 & 12&7   & 0.3   \\
   \uspace0.5 \dspace
                  & $J$=3--2 \see{e}                & 345795.975
                  & 2.37  & 130 &  126.1 & 3.7 &  9&66  & 0.08  \\
   \hline
   \end{tabular}
   \end{flushleft}
  {\footnotesize \begin{enumerate} \renewcommand{\labelenumi}{\alph{enumi})}
   \item For non-detections, the single channel 3$\sigma $ noise level is given,
         divided by the square root of the number of expected line channels.
   \item r.m.s noise for an individual channel.
   \item The intensities are integrated from \vLSR\ = 122 to 130\,\kms.
         For the detected lines, errors are one standard deviation.
         For undetected transitions, upper 3$\sigma$ limits are given.
   \item Tentative detection
   \item The spectrum has been convolved to the beam size
         of the $J$=1--0 transition ($\approx$ 43\arcsec).
  \end{enumerate} }
\end{table*}
\normalsize

\section{Results}
 \label{sec:MC3-Results}

   Fig.\,\ref{fig:contour-LIRS36} shows maps of the LIRS\,36 complex,
   obtained in \twCO\ $J$=1--0, 2--1, 3--2, and \thCO\ $J$=1--0 with spacings of
   28\arcsec, 10\arcsec, 10\arcsec, and 28\arcsec\ for the central region,
   respectively.
   The 1--0 map does not spatially resolve the source.
   While the 2--1 line data are marginally resolving
   the source in E--W direction (full width to half power (FWHP) size
   $\approx$ 40\arcsec, corresponding to 12\,pc at $D$ = 60\,kpc), the
   emitting region is elongated along the N-S axis
   (FWHP size $\approx$ 60\arcsec; 17\,pc at $D$ = 60\,kpc).
   The $J$=3--2 emission shows a similar but more compact distribution.
   Radial velocities are slightly above
   126\,\kms\ toward the central and western parts of the complex and
   124 -- 126\,\kms\ at southern, eastern and northern offsets.

   A total of 12 molecular species has been observed.
   We have detected nine of these in a total of 18 rotational transitions,
   including isotopic lines.
   Two tentative detections have also been obtained.
   This demonstrates that it is possible to carry out molecular multiline
   studies, including many species, for selected regions of the SMC,
   extending the range of metallicities by more than a factor of two
   below that of the Large Magellanic Cloud (cf.\ Westerlund 1990).
   Spectra and line parameters obtained by gaussian fits are displayed
   in Fig.\,\ref{fig:LIRS36} and Table\,\ref{tbl:LIRS36}.
   The three \FORM\ $J$=2--1 $K_{\rm a}$ = 0,1 transitions have been measured,
   while the $J_{K_{\rm a}, K_{\rm c}}$ = 3$_{0,3}$--2$_{0,2}$ transition is
   only tentatively detected.
   Three rotational transitions of CS, the $J$=2--1, 3--2, and 5--4 lines,
   have been measured in the main species and two in \CfoS.
   While the \CfoS(3--2) line is apparently detected, we find no evidence
   for the corresponding 2--1 transition.
   Although studies of Galactic star-forming regions (Chin \etal\ 1996) imply
   that CS (3--2) transitions from rare isotopes are more easily detected than
   the corresponding 2--1 lines, we classify our \CfoS\ detection as tentative.
   The non-detection of the \HCOp\ and HCN $J$=3--2 transitions is not
   surprising in view of the relatively high upper limits obtained
   in the warm humid weather conditions of the Chilean summer.

   Among the rare isotopic species of CO, \thCO\ is seen but not \CeiO.
   The tentative detection of \CCH\ (Chin \etal\ 1997) is confirmed;
   the relative line intensities of its two observed hyperfine components imply
   that the emission is optically thin (cf.\ Nyman 1984).
   CN remains unconfirmed, in spite of the higher sensitivity
   of the data presented here.

\section{Discussion}
 \label{sec:MC3-Discussion}

   LIRS\,36 is one of the most prominent far infrared sources of the SMC
   (Schwering \& Israel 1989).
   It is an IRAS point source with a cool dust spectrum (\Tdust\
   $\approx$ 30\,K) and an infrared luminosity of $3 \times 10^5$\,\solum\
   (we applied the method outlined by Wouterloot \& Walmsley 1986;
   $D$ = 60\.kpc).

\subsection{The X-factor}
 \label{sec:MC3-X-factor}

   Assuming that the interstellar medium is in virial equilibrium
   at all linear scales and that the line width is representing the
   cloud's velocity distribution, we can deduce the
   $X$ = $N_{\rm H_2}$/$I_{\rm CO}$ conversion factor from
$$
     \Mvir\ = 190\ R \ (\Delta v_{1/2})^2,
$$
$$
     \Mgas\ = 1.36\,\,\left[N_{\rm H_2}/I_{\rm CO}\right]\ m_{\rm H_2}\
              \sum_{i=1}^n \frac{I_{\rm u,i} + I_{\rm l,i}}2 \ A_{\rm i},
$$
   and
$$
     \Mvir\ = \Mgas\ .
$$
   \Mvir\ is the virial mass in solar units (\Msol), \Mgas\ is the cloud mass
   derived from $I_{\rm CO}$ (also in solar units),
   $R$ is the cloud radius in pc,
   \delv\ is the total linewidth in \kms,
   1.36 is the correction to include helium and metals,
   $N_{\rm H_2}$ is the \MOLH\ column density in \cmsq,
   $I_{\rm CO}$ denotes the integrated CO (1--0) line intensity in \Kkms,
   $m_{\rm H_2}$ is the mass of an H$_2$ molecule,
   and $I_{\rm u,i}$ and $I_{\rm l,i}$ (in \Kkms) are the integrated
   intensities of the upper and lower contours confining the area
   $A_{\rm i}$ (in cm$^2$; $n$: number of contours).
   The factor 190 in the virial mass equation refers to a $1/r$ density
   distribution and is consistent, within a factor of two, with constant
   and $1/r^2$ density profiles (MacLaren \etal\ 1988).
   Making use of the beam deconvolved FWHP cloud size deduced from the 2--1
   spectra (cf.\ Sect.\,\ref{sec:MC3-Results}) and
   accounting for all the CO emission from the cloud,
   we obtain $X_{\rm LIRS36}$ = $4.8 \times 10^{20}$\,\cmsq\,(\Kkms)$^{-1}$.
   This is a factor of two larger than the local Galactic disk value and leads
   to a \MOLH\ column density (averaged over the map) of $N$(\MOLH) $\approx$
   $3.7 \times 10^{21}$\,\cmsq, that is consistent with the characteristic
   column density of nearby Galactic molecular clouds (\eg\ Larson 1981).
   The cloud follows the correlations between \delv, mass, and size,
   proposed by Larson (1981), and the average number density
   becomes \numd\ $\approx$ 100\,\percc.

   Analysing CO clouds with a wide range of radii,
   Rubio \etal\ (1993a) obtained $X_{\rm SMC}$ $\approx$
   $9 \times 10^{20}$\,($R$/10pc)$^{0.7}$ \cmsq\,(\Kkms)$^{-1}$.
   The correlation between $X$-factor and linear scale $R$ was interpreted
   in terms of an increased rate of photodissociation of CO due to a strong UV
   radiation field and a low gas to dust mass ratio and carbon abundance.
   Moreover, Rubio \etal\ (1993a) suggest a predominantly
   atomic intercloud medium.
   Our observations trace the SMC with a characteristic linear scale of
   10\,pc that is slightly smaller than the size of the molecular complex;
   our LIRS\,36 $X$-factor is half of the value obtained
   by Rubio \etal\ (1993a) from their sample of clouds.

   Specifically for LIRS\,36, Rubio \etal\ (1993b) find a virial mass
   of $5.1 \times 10^4$\,\Msol.
   The mass derived by us is $1.6 \times 10^4$\,\Msol.
   The application of a slightly smaller linewidth (3.6\,\kms) than those
   of Rubio \etal\ (1993b) and Lequeux \etal\ (1994) (3.8\,\kms) can be
   justified by the good agreement of the \delv\ derived from our \twCO(1--0),
   (2--1), (3--2) and \thCO(1--0) observations (Table \ref{tbl:LIRS36}).
   Since the difference caused by linewidths is negligible ($\approx$ 10\%),
   the discrepancy in \Mvir\ is caused mainly by the cloud radius $R$,
   18.6 versus 9.8\,pc.
   The discrepancy in $R$ is caused by two effects:
   Firstly, Rubio \etal\ (1993a) use the CO non-deconvolved
   0.4\,\Kkms\ contour (in units of antenna temperature) as the cloud boundary;
   we use instead the deconvolved FWHP contour.
   While Rubio \etal\ (1993a) cloud sizes are based on CO (1--0) data,
   our cloud size is based on the CO (2--1) transition,
   assuming that 1--0 and 2--1 emission have a similar extent.
   Since CO (1--0) and (2--1) cloud sizes should be similar for the SMC
   (cf.\ Lequeux \etal\ 1994) and since cloud size measurements
   based on higher resolution CO (2--1) spectra provide more accurate
   results, our smaller $R$ value should be preferred.
   The determination of $R$ (defined by the FWHP contour) from CO (1--0),
   not (2--1) line spectra, may lead to a {\it systematic overestimate} of
   the average $X$-value for clouds with radii $R$ $\approx$ 10\,pc.
   A small $X$-factor may imply that an atomic interclump medium
   can only play a minor role in the molecular cores.
   Detailed studies of additional SMC cores are needed to demonstrate that small
   scale SMC $X$-factors are generally as low as is indicated by our data.

\subsection{Molecular abundances}
 \label{sec:MC3-abundances}

\subsubsection{$^{12}$CO versus $^{13}$CO}
 \label{sec:MC3-13CO}

   Our measured \twCO\ and \thCO\ line intensities are larger
   than those reported by Rubio \etal\ (1993b):
   While the discrepancy is typically 20\%, our \thCO(2--1) peak
   line temperature is higher by a factor of two.
   This is partially compensated by the exceptionally large linewidth assigned
   to this line by Rubio \etal\ (1993b), so that the ratio of integrated line
   intensities only becomes 1.3.
   Our beam size corrected $J$=2--1/$J$=1--0 \twCO\ and \thCO\
   line intensity ratios are $1.21 \pm 0.03$ and $1.50 \pm 0.09$
   (the 1$\sigma$ error refers to the noise in the individual spectra
   and does not include calibration uncertainties
   that are given in Sect.\,\ref{sec:MC3-Observations}).
   This should be compared with 1.16 and 1.25 from Rubio \etal\ (1993b).
   Most of the discrepancies in line intensity ratios can be explained
   in terms of the higher sensitivity of our $\lambda$ = 1.3\,mm data and
   the smaller 1.3\,mm SEST beam efficiency adopted by us.

   The integrated line intensity ratios from the presumably optically thin
   \thCO\ transitions are consistent with a density of
   \numd\ $\approx$ $10^4$\,\percc\ (for \Tkin\ $\approx$ 20\,K,
   applying a Large Velocity Gradient (LVG) radiative transfer code;
   see also Lequeux \etal\ 1994).
   Since the density is higher than the virial density obtained
   in Sect.\,\ref{sec:MC3-X-factor}, there must be small scale structure
   that is not resolved by our $\ga$ 15\arcsec\ beam.

\subsubsection{$^{13}$CO versus C$^{18}$O}
 \label{sec:MC3-C18O}

   The non-detection of \CeiO\ is consistent with the large \thCO/\CeiO\
   line intensity ratios observed toward the LMC
   (Johansson \etal\ 1994; Chin \etal\ 1997).
   We find $I$(\thCO)/$I$(\CeiO) $>$ 35 (a 3$\sigma$ limit).
   While the limit is less stringent than those obtained from star-forming
   regions of the LMC, the actual value must be larger than
   the characteristic Galactic line intensity ratio,
   $I$(\thCO)/$I$(\CeiO) $\approx$ 10 (cf.\ Lada 1976; Johansson \etal\ 1994).
   Whether this is caused by `anomalies' in the isotopic abundances
   relative to those of the Galactic disk (cf.\ Henkel \& Mauersberger 1993),
   whether it is caused by isotope selective photodissociation
   (\eg\ van Dishoeck \& Black 1988; Fuente \etal\ 1993) or
   whether it is caused by fractionation in a partially ionized medium
   (cf.\ Watson \etal\ 1976) remains to be seen.
   The small \CeiO/\CseO\ line intensity ratio in the LMC
   (Johansson \etal\ 1994) hints at a low \O18 \ abundance,
   but we do not know whether this also holds for the SMC.
   $I$(\thCO)/$I$(\CeiO) $\approx$ 40, as measured toward the Galactic
   \HII\ region S\,68 (Bally \& Langer 1982), demontrates that fractionation
   can, in principle, account for the observed line intensity ratio anomaly.
   Assuming that both \thCO\ and \CeiO\ are optically thin and have the
   same excitation temperature, that the \thCO\ abundance is enhanced by
   the maximum factor permitted by chemical fractionation,
   e$^{35\,{\rm K/T_{\rm kin}}}$, and that the \C13 /\C12 \ : \O18 /\O16 \
   ratio is $\approx$ 7 as in the local Galactic disk
   (\eg\ Wilson \& Rood 1994), we obtain for
   $I$(\thCO)/$I$(\CeiO) $\approx$ 40 with
$$
     {\rm e}^{35\,{\rm K}/T_{\rm kin}} =
     \left[I(^{13}{\rm CO}/I({\rm C}^{18}{\rm O}\right] /
     \left[^{13}{\rm C}/^{12}{\rm C}\,\, : \,\,^{18}{\rm O}/^{16}{\rm O}\right]
$$
   a kinetic temperature of $\approx$ 20\,K.
   The temperature is consistent with the cloud core model temperatures
   suggested by Lequeux \etal\ (1994).
   A large \thCO\ abundance enhancement caused by fractionation is plausible
   in the interstellar medium of the SMC with
   its small dust opacities and strong UV radiation field.

\subsubsection{\HCOp, HCN, and HNC}
 \label{sec:MC3-HCO}

   The relative $J$=1--0 intensities of \HCOp, HCN, and HNC, three molecules
   with similar rotational constants and electric dipole moments, also follow
   the trend obtained toward star-forming regions of the LMC.
   For the $J$=1--0 transition, $I$(\HCOp)/$I$(HCN) and $I$(HCN)/$I$(HNC) $>$ 1.
   This can be interpreted in terms of warm molecular gas,
   coupled with an intense ionization flux from supernovae,
   and with \HCOp\ arising from a larger volume than HCN
   (cf.\ Johansson \etal\ 1994; Chin \etal\ 1997).

\subsubsection{CN}
 \label{sec:MC3-CN}

   The most unexpected result of our line survey is
   the absence of a detectable CN signal.
   CS $J$=2--1 and 3--2 lines tend to be weaker than those of CN
   in most Galactic and extragalactic sources
   (cf.\ Henkel \etal\ 1988, 1990, 1993; Mauersberger \& Henkel 1989;
   Nyman \& Millar 1989; Ziurys \etal\ 1989; Nyman \etal\ 1993).
   Toward LIRS\,36, CN remains undetected but we may have seen
   a rare isotopic species of CS.

\subsubsection{CS and the cloud density}
 \label{sec:MC3-CS}

   At least three of the five observed CS transitions were detected
   (see Table \ref{tbl:LIRS36} and Fig.\,\ref{fig:LIRS36}).
   In the case that the CS emitting region is as extended
   as the CO emitting region, we have to apply the CO 1.3\,mm line intensity
   correction factor, 0.7, for CS $J$=5--4.
   This factor can be extracted from Table\,\ref{tbl:LIRS36}
   and was also recommended by Rubio \etal\ (1993b) to account for
   the small 1.3\,mm beam size when comparing 1.3\, with 3\,mm data.
   Interpolating, 0.85 is then the appropriate correction factor
   for our $\lambda $ = 2\,mm data.
   If CS arises instead from a point source, the beam size corrections
   become 0.25 at 1.3\,mm and 0.5 at 2\,mm wavelength.
   Since CS requires higher densities than CO to become detectable and
   since CS optical depths tend to be smaller (cf.\ Bohlin \etal\ 1978;
   Linke \& Goldsmith 1980; Larson 1981; Bachiller \& Cernicharo 1986),
   actual correction factors will be 0.25 $\leq$ $f_{1.3}$ $\leq$ 0.7 at 1.3\,mm
   and 0.5 $\leq$ $f_2$ $\leq$ 0.85 at 2\,mm.
   Applying an LVG model describing a spherical or planparallel cloud
   with uniform density and temperature (for details, see Mauersberger
   \& Henkel 1989), we then find that in the optically thin case
   with \Tkin\ = 20\,K (cf.\ Lequeux \etal\ 1994) the number density becomes
   10$^{6.0}$\,\percc\ $\leq$ \numd\ $\leq$ 10$^{6.6}$\,\percc.
   Also accounting for the standard deviation of the
   integrated CS (5--4) line intensity (Table\,\ref{tbl:LIRS36}),
   we find \numd\ = 10$^{5.9 - 6.9}$\,\percc.
   With a high 1.3\,mm beam efficiency of 0.6 (0.45 is used throughout this
   paper; see Sect.\,\ref{sec:MC3-Observations})
   we would get instead \numd\ = 10$^{5.7 - 6.4}$\,\percc.
   In order to fully explore the range of relevant kinetic temperatures,
   we have also calculated densities for \Tkin\ = 100\,K, a very high
   value (see \eg\ Mauersberger \etal\ 1990; Lequeux \etal\ 1994).
   Accounting again for the standard deviation in the integrated
   CS (5--4) line intensity,
   we now find \numd\ = 10$^{5.2 - 5.7}$\,\percc.
   With all uncertainties considered, we thus conclude that
   densities must be high, of order $10^5$ to $10^7$\,\percc.
   This is consistent with the density derived from \FORM\
   if the $J_{Ka,Kc}$ = 3$_{0,3} - 2_{0,2}$ transition is detected and if
   $\tau_{\rm H_2CO, J=2-1}$ $\la$ 1.
   The densities are much higher than those derived from \twCO\
   in Sect.\,\ref{sec:MC3-X-factor} and from \thCO\ in Sect.\,\ref{sec:MC3-13CO}
   and are mainly based on a comparison of 2--1 with 5--4 CS lines.
   Because of their relatively large separation in excitation energy,
   they are much more sensitive density tracers than
   \eg\ the 2--1 and 3--2 transitions.

   So far, we have assumed that CS lines are optically thin,
   which was suggested by Johansson \etal\ (1994) for N\,159,
   an LMC \HII\ region that exhibits stronger line emission than LIRS\,36.
   With a \CtwS/\CfoS\ abundance ratio of $\approx$ 20 in the disk of
   the Milky Way however (see Chin \etal\ 1996), a comparison of
   \CtwS\ and \CfoS\ (3--2) line intensities (Table\,\ref{tbl:LIRS36})
   leads to an optical depth of $\approx$ 5 for the \CtwS\ (3--2) line.
   There are hence three possibilities:
   (1) The CS lines are moderately optically thick ($\tau$ $\la$ 5).
       With \numd\ $>$ 10$^{4.5}$\,\percc, this does not imply a
       significant reduction of the densities estimated above.
   (2) The CS lines are optically thin and the $^{32}$S/$^{34}$S ratio
       drastically deviates from that of the Galactic disk.
       This leads to a \S32 /\S34 \ isotope ratio of $\approx$ 5,
       implying that in the SMC massive stars and type Ia supernovae
       provide an isotopic mixture that is quite different from
       that of the more metal rich disk of the Milky Way.
   (3) Our tentative \CfoS\ (3--2) detection, although appearing to be
       quite convincing, is not real.
   Another independent measurement of \CfoS\
   is required to discriminate between these possibilities.

\begin{table}
   \caption[]{Molecular column densities}
 \label{tb2:LIRS36}
 \scriptsize
   \begin{flushleft}
   \begin{tabular}{l c c r@{~$\times$~}l r@{~$\times$~}l}
%
   \hline
   \uspace1.5
   Molecule       & \Tkin & \numd
      & \multicolumn{2}{c}{Column density}
      & \multicolumn{2}{c}{Fractional abundance} \\
   \uspace0.5 \dspace
            & [K]   & [\percc]
      & \multicolumn{2}{c}{[\cmsq]}
      & \multicolumn{2}{c}{~} \\
   \hline
   \uspace1
   \MOLH\ \see{a} & --- &    ---     & ~~~~~3 & 10$^{21}$
                                              & \multicolumn{2}{c}{1}       \\
   \uspace0.5
   \twCO\ \see{b} & --- &    ---     &    3 & 10$^{16}$
                                              & \hspace*{8mm}1 & 10$^{-5}$  \\
   \uspace0.5
   \thCO\ \see{b} &  20 & 10$^{3.9}$ &    7 & 10$^{14}$ &    2 & 10$^{-7}$  \\
   \uspace0.5
   CN \see{c}     & --- &    ---     &$<$ 2 & 10$^{12}$ &$<$ 7 & 10$^{-10}$ \\
   \uspace0.5
   CS \see{d}     &  20 & 10$^{6.5}$ &    1 & 10$^{12}$ &    3 & 10$^{-10}$ \\
   \uspace0.5
   \HCOp\ \see{e} &  30 & 10$^{5.0}$ &    7 & 10$^{11}$ &    2 & 10$^{-10}$ \\
   \uspace0.5
   HCN \see{f}    &  20 & 10$^{6.5}$ &    2 & 10$^{11}$ &    7 & 10$^{-11}$ \\
   \uspace0.5
   HNC \see{g}    &  20 & 10$^{5.5}$ &    9 & 10$^{10}$ &    3 & 10$^{-11}$ \\
   \uspace0.5 \dspace
   \FORM\ \see{h} &  20 & 10$^{5.5}$ &    9 & 10$^{11}$ &    3 & 10$^{-10}$ \\
   \hline
   \end{tabular}
   \end{flushleft}
   {\footnotesize \begin{enumerate} \renewcommand{\labelenumi}{\alph{enumi})}
     \item Column density estimated from the virial mass
           (see Sect.\,\ref{sec:MC3-X-factor}).
     \item The column density has been derived from an LVG simulation,
           reproducing the beam size corrected \thCO\ $J$=1--0, 2--1,
           and 3--2 line intensities (Table\,\ref{tbl:LIRS36} and
           Sect.\,\ref{sec:MC3-13CO}) and applying a velocity gradient
           of \delv/2$R$ = 0.27\,\kms\,pc$^{-1}$.
           With a \twCO/\thCO\ abundance ratio of 50 (Johansson \etal\ 1994)
           the CO column density becomes $3.5 \times 10^{16}$\,\cmsq.
           The fractional abundance is [CO]/[\MOLH] = $10^{-5}$.
           This is $\approx$ 20\% of the value in Galactic clouds
           (\eg\ Wilson \etal\ 1986), reflecting the
           low carbon abundance in star-forming regions of the SMC
           ([C/H] $\approx$ $-$1.3; Westerlund 1990).
     \item CN excitation temperatures are small ($<$ 10\,K) at low densities
           (\numd\ $<$ $10^4$ \percc), while abundances decrease at higher
           densities (\eg\ Turner \& Gammon 1975; Churchwell 1980;
           Churchwell \& Bieging 1982, 1983; Crutcher \etal\ 1984;
           Henkel \etal\ 1988; Johansson \etal\ 1994).
           With an excitation temperature of 10\,K and accounting for the fact
           that the LTE intensity of the $N$=1--0 $J$=3/2--1/2 $F$=5/2--3/2 CN
           line only comprises one third of the total $N$=1--0 line intensity
           under optically thin conditions (Skatrud \etal\ 1983), we find the
           given beam averaged 3$\sigma$ upper CN column density limit.
           See also Sect.\,\ref{sec:MC3-CN}.
     \item See Sect.\,\ref{sec:MC3-CS}.
     \item The column density is derived in the same way as for HCN
           (see footnote f).
           Different densities and temperatures were chosen to account for
           \HCOp\ depletion at highest densities and for the suspected large
           spatial extent of the \HCOp\ emission region (\eg\ Chin \etal\ 1997).
     \item HCN is suspected to emit mainly from the coolest and densest parts
	   of the cloud (cf.\ Johansson \etal\ 1994; Chin \etal\ 1997), so that
           \Tkin\ = 20\,K and \numd\ = 10$^{6.5}$\,\percc\ are appropriate.
           The column density was derived with an LVG code
           (see also footnote b).
           The HCN (3--2) upper limit (Table\,\ref{tbl:LIRS36}) is
           not small enough to provide a stringent density limit.
     \item The column density is derived in the same way as for HCN
           (see footnote f).
           Since [HCN]/[HNC] abundance ratios tend to drop with rising density
           (Schilke \etal\ 1992), \numd\ = 10$^{5.5}$\,\percc\ is justified.
     \item The column density was derived from an LVG simulation with $J$=3--2
           line intensities being slightly weaker (by $\approx$ 30\%) than
           the 2--1 lines (see also Sect.\,\ref{sec:MC3-CS}).
   \end{enumerate} }

\end{table}
\normalsize

\subsubsection{Column densities}
 \label{sec:MC3-densities}

   Having estimated the molecular densities in Sects.\,\ref{sec:MC3-X-factor},
   \ref{sec:MC3-13CO}, and \ref{sec:MC3-CS} and having a quantitative idea
   of the kinetic temperature distribution (Lequeux \etal\ 1994), we can now
   also determine the column densities for various molecular species.
   Both temperature and density vary inside the cloud and each molecule
   and transition is tracing gas at different temperature and density;
   error limits are thus difficult to assess.
   Nevertheless, our data allow a rough estimate of molecular abundances
   in the optically thin limit.
   This is the appropriate approach, firstly because we do not know the
   optical depth of most lines and secondly because Johansson \etal\ (1994)
   find \HCOp, HCN, and CS to be (almost) optically thin in their mm-wave
   transitions toward the prominent LMC star-forming region N\,159.
   Assumed kinetic temperatures, densities, and resulting absolute
   and fractional (relative to \MOLH) column densities for a
   $\approx$ 50\arcsec\ (15\,pc) beam size are given in Table\,\ref{tb2:LIRS36}.

   A comparison with Galactic disk values directly shows that
   the LIRS\,36 fractional abundances, [X]/[\MOLH],
   are much smaller (\eg\ Blake \etal\ 1987):
   For CO, CS, \HCOp, and \FORM, underabundances relative to the Orion extended
   ridge or to the prototypical dark cloud TMC-1 are $\approx$ 10.
   For HCN and HNC and probably for CN as well, underabundances reach
   two orders of magnitude and are thus even surpassing those reported
   by Johansson (1997) for the LMC star-forming region N\,159.
   It is quite remarkable that CN is not enhanced relative
   to HCN and HNC as observed in Galactic disk photodissociation regions
   with high UV fluxes (Fuente \etal\ 1993; Greaves \& Church 1996).
   We may suspect that the low LIRS\,36 fractional abundances of
   HCN, HNC, and CN are related to a small nitrogen abundance.
   With average values of [N/H] = $-$1.1 and [C/H] = $-$1.3 in
   star-forming regions of the SMC (Westerlund 1990), however,
   nitrogen is not the most depleted among the CNO elements.
   It thus remains to be seen whether the HCN, HNC, and CN fractional abundances
   are related to an exceptionally small nitrogen abundance in LIRS\,36
   or whether they are caused by a chemical process that is found
   in other molecular SMC cores as well.

 \begin{figure*}[tbh]
    \hspace*{10 mm}
    \psfig{figure=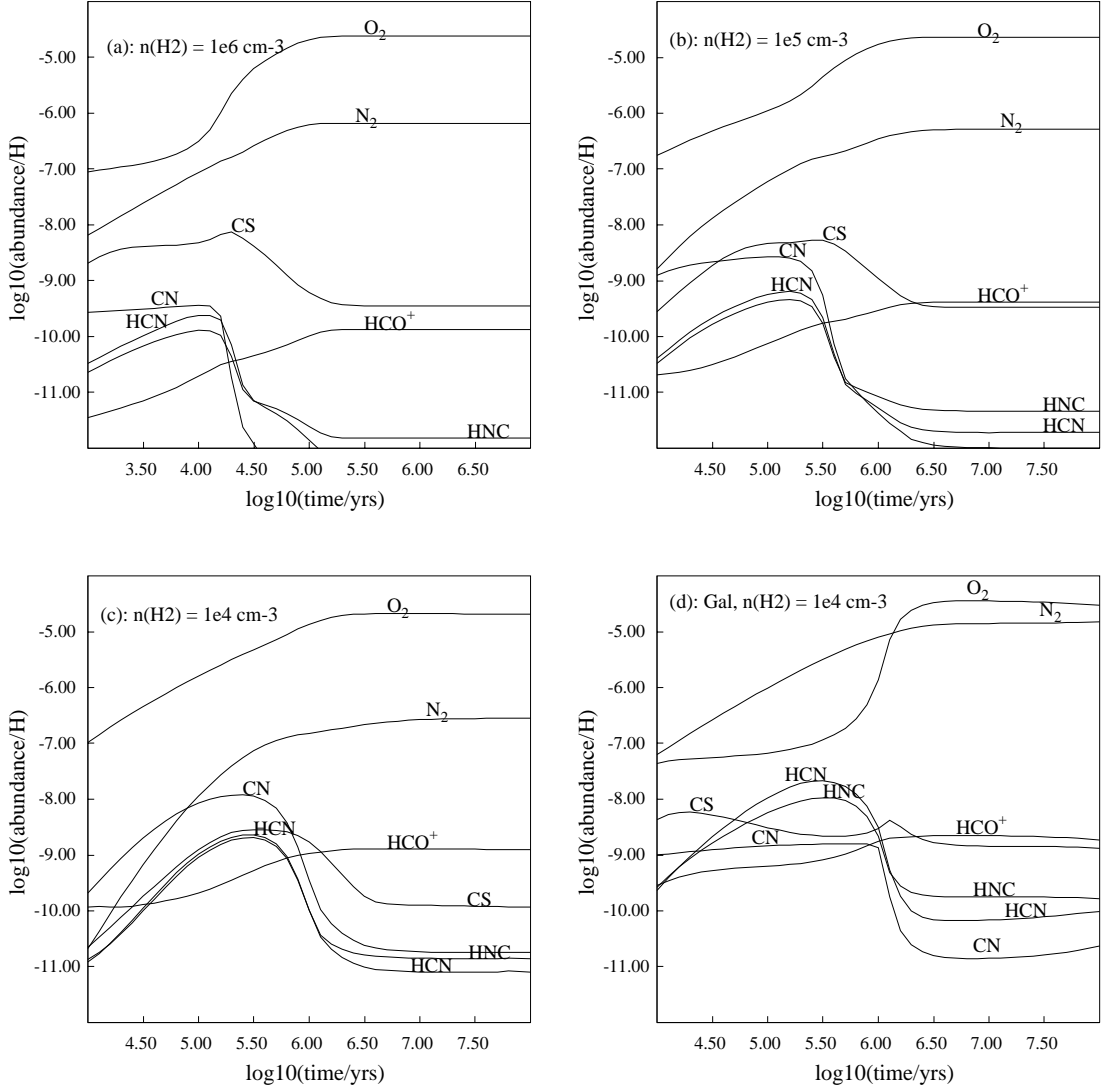,height=15cm}
    \vspace*{-5 mm}
    \caption[]
            {Chemical abundances of selected molecules:
             SMC molecular abundances for densities \numd\ of
             (a) 10$^4$, (b) 10$^5$, and (c) 10$^6$\,\percc\ as a function of
             chemical timecscale, using elemental abundances from
             the `S2' model of Millar \& Herbst (1990);
             (d) `Galactic' model (model `G' of Millar \& Herbst 1990).}
  \label{fig:Millar-model}
 \end{figure*}

\subsubsection{Chemical model calculations}
 \label{sec:MC3-model}

   We have used the models described by Chin \etal\ (1997) to investigate
   the chemical evolution of LIRS\,36 using elemental abundances of C, N, and O
   from the S2 model of Millar \& Herbst (1990).
   Because observed emission clearly arises in regions having different
   densities and because we have no information on the density profile within
   the molecular cloud nor on the the size of the high density CS clump,
   we have considered a simple constant density model with densities in
   the range \numd\ = $10^4 - 10^6$\,\percc\ and \Tkin\ = 20\,K.
   For comparison, we have also calculated a `Galactic' model
   (model `G' of Millar \& Herbst 1990) for
   \numd\ = $10^4$\,\percc\ (Fig.\,\ref{fig:Millar-model}).
   In addition to a lower metallicity, the SMC models have a smaller
   dust-to-gas ratio, and hence H$_2$ formation rate, by a ratio of 6.5 and
   a UV intensity larger by a factor of 4, compared to Galactic values.
   There are conflicting results on the appropriate cosmic-ray ionisation rate.
   Chi \& Wolfendale (1993) have argued that, based on gamma-ray observations,
   the rate in the SMC is at most 11\% of the Galactic value.
   However, observations of \HCOp\ in the LMC and SMC (Chin \etal\ 1997)
   indicate that the rate is larger than that implied by Chi \& Wolfendale.
   In these models we use a rate of $5.2 \times 10^{-17}$\,s$^{-1}$,
   three times larger than the Galactic value.
   The main impact of this rate is to increase the ionisation fraction
   and to reduce the chemical time-scales.
   Finally, we have used a visual extinction of 10 mag.\
   in calculating photorates.
   While this is obviously too large for the molecular envelope of LIRS\,36,
   the central region, in which densities greater than $10^5$\,\percc\
   are derived from CS, could easily have an extinction large enough
   to exclude photoeffects even with a low dust-to-gas ratio.

   The results of our four model calculations are summarised in
   Fig.\,\ref{fig:Millar-model} and indicate that the abundance of CN
   is sensitive to the O$_2$ abundance since this species destroys CN with
   a fast, measured rate coefficient of $2.4 \times 10^{-11} \cdot$
   (\Tkin/300\,K)$^{-0.6}$\,cm$^3$\,s$^{-1}$ down to 20\,K (Sims \etal\ 1994).
   The fractional abundance of O$_2$ increases as one goes from the Galactic
   to the SMC model at \numd\ = $10^4$\,\percc\, and as density increases
   because of the decreasing abundance of C atoms which destroy O$_2$.
   In model S2 there is more free oxygen remaining after
   all available carbon has been tied up in CO so that the importance
   of the destruction of CN by O$_2$ increases.
   Note that the $\bar{O}/\bar{C}$ ratio, where $\bar{X}$ represents the
   elemental abundance of species $X$, varies between 2.4 (model G) and
   10.5 (model S2) while the difference between the elemental abundances
   of $\bar{O}$ and $\bar{C}$ is $2.1 \times 10^{-4}$ (model G) and
   $5.7 \times 10^{-5}$ (model S2), and the ratio
   ($\bar{O} - \bar{C})/\bar{C}$, increases from 1.4 (model G)
   to 10.4 (model S2).
   These values show that in model S2 there is a lot of oxygen
   remaining after all the carbon has been processed into CO and thus the
   importance of CN destruction by O$_2$ is more important in the SMC models.

   The fractional abundance of CN decreases by about a factor of 3
   as one goes from model G to S2 and by a further factor of 100 as
   the density increases from $10^4$ to $10^6$\,\percc.
   As a result, the abundance of CN becomes less than or comparable with
   those of HCN and HNC at high density.
   On the other hand, CS is not destroyed by O$_2$ or O atoms and its abundance
   changes by less than a factor of two over all the models investigated.
   Note that the maximum fractional abundance of CN at 10$^6$\,\percc\
   is only $4 \times 10^{-10}$ for chemical timescales $<$ $10^5$\,yrs;
   for longer times, the abundance falls rapidly to
   its steady-state value of $4 \times 10^{-14}$.

\section{Conclusions}
 \label{sec:MC3-Conclusions}

   Having made a mm-wave molecular survey of the LIRS\,36 star-forming region
   in the SMC, we obtain the following main results:

\begin{enumerate} \renewcommand{\labelenumi}{(\arabic{enumi})}
   \item We have detected CO, CS, SO, \CCH, \HCOp, HCN, HNC, \FORM, and \CYCP.
         NO, \METH, and, surprisingly, CN were not detected.
         Among the rare isotopic species, we have seen \thCO\ and
         tentatively C$^{34}$S, but not \CeiO.

   \item For a characteristical scale length of $\approx$ 10\,pc and
         assuming virial equilibrium,
         the $X$ = $N$(\MOLH)/$I_{\rm CO}$ conversion factor is with
         $\approx$ $4.8 \times 10^{20}$\,\cmsq\,(\Kkms)$^{-1}$ a factor of two
         larger than the local Galactic disk $X$-factor.
         SMC $X$-factors given by Rubio \etal\ (1993a) for this scale length
         may have to be reduced by half an order of magnitude.

   \item Density estimates range from \numd\ $\approx$ 100\,\percc\ (deduced
         from the virial masss and the spatial extent of the CO emission) over
         $10^4$\,\percc\ (from \thCO) to $10^5$ -- $10^7$\,\percc\ (from CS).
         The observed CS transitions provide strong
         evidence for the presence of a very dense core with a density
         likely surpassing $10^6$\,\percc.

   \item As in the LMC cloud cores, the $I$(\thCO)/$I$(\CeiO) line intensity
         ratio is larger than the usual values encountered in the Galactic disk.
         Whether this is caused by isotopic abundance anomalies,
         by isotope selective photodissociation, or
         by chemical fractionation remains an open question.

   \item $I$(\HCOp)/$I$(HCN) and $I$(HCN)/$I$(HNC) line intensity ratios
         are $>$ 1, consistent with molecular emission from a warm (\Tkin\ $>$
         10\,K) molecular environment exposed to a high ionizing flux.

   \item Fractional (relative) abundances of CO, CS, \HCOp, and \FORM\
         are an order of magnitude below those of the Galactic disk.
         HCN, HNC, and likely also CN are even underabundant
         by two orders of magnitude.
         Whether this is reflecting a particularly small nitrogen abundance
         in LIRS\,36 or whether this is a common chemical peculiarity
         of other SMC cores as well remains open.
         The non-detection of CN can be explained in terms of the high cloud
         density and a (relatively) high fractional abundance of
         O$_2$ that is destroying CN but not CS.
\end{enumerate}

\acknowledgements{
   We thank A.~Heikkil\"a for critically reading the manuscript.
   YNC thanks for financial support through National Science Concil
   of Taiwan grant \hbox{86-2112-M001-032}.
   TJM is supported by a grant from PPARC.
   Financial support enabling JBW to travel to SEST was provided by
   the Department of Industry Science and Tourism, Australia.
}


\begin{thebibliography}{}
 \bibitem[1999]{xyz}
   Bachiller R., Cernicharo J., 1986, \AaA\ \vol{166}, 283
 \bibitem[1999]{xyz}
   Bally J., Langer W.D., 1982, \ApJ\ \vol{255}, 143
 \bibitem[1999]{xyz}
   Blake G.A., Sutton E.C., Masson C.R., Phillips T.G., 1987,
     \ApJ\ \vol{315}, 621
 \bibitem[1999]{xyz}
   Bohlin R.C., Savage B.D., Drake J.F., 1978, \ApJ\ \vol{224}, 132
 \bibitem[1999]{xyz}
   Chi X., Wolfendale A.W., 1993, J. Phys. G, \vol{19}, 795
 \bibitem[1999]{xyz}
   Chin Y.-N., Henkel C., Whiteoak J.B., Langer N., Churchwell E.B.,
     1996, \AaA\ \vol{305}, 960
 \bibitem[1999]{xyz}
   Chin Y.-N., Henkel C., Whiteoak J.B., \etal, 1997, \AaA\ \vol{317}, 548
 \bibitem[1999]{xyz}
   Churchwell E., 1980, \ApJ\ \vol{240}, 811
 \bibitem[1999]{xyz}
   Churchwell E., Bieging J.H., 1982, \ApJ\ \vol{258}, 515
 \bibitem[1999]{xyz}
   Churchwell E., Bieging J.H., 1983, \ApJ\ \vol{265}, 216
 \bibitem[1999]{xyz}
   Crutcher R.M., Churchwell E., Ziurys L.M., 1984, \ApJ\ \vol{283}, 668
 \bibitem[1999]{xyz}
   Fuente A., Mart\'in-Pintado J., Cernicharo J., Bachiller R., 1993,
     \AaA\ \vol{276}, 473
 \bibitem[1999]{xyz}
   Greaves J.S., Church S.E., 1996, \MNRAS\ \vol{283}, 1179
 \bibitem[1999]{xyz}
   Heikkil\"a A., Johansson L.E.B., Olofsson H., 1997, IAU Symp. 178 Abstract
     Book, Molecules in Astrophysics, Probes and Processess,
     eds.\ D.J.~Jansen, M.R.~Hogerheijde, E.F.~van Dishoeck, Leiden, p301
 \bibitem[1999]{xyz}
   Henkel C., Mauersberger R., 1993, \AaA\ \vol{274}, 730
 \bibitem[1999]{xyz}
   Henkel C., Mauersberger R., Schilke P., 1988, \AaA\ \vol{201}, L23
 \bibitem[1999]{xyz}
   Henkel C., Whiteoak J.B., Nyman L.-\AA, Harju J., 1990, \AaA\ \vol{230}, L5
 \bibitem[1999]{xyz}
   Henkel C., Mauersberger R., Wiklind T., H\"uttemeister S., Lemme C,
     Millar T.J., 1993, \AaA\ \vol{268}, L17
 \bibitem[1999]{xyz}
   Johansson L.E.B., 1991, IAU Symp. 146, Dynamics of Galaxies and their
     Molecular Cloud Distributions, eds.\ F.~Combes, F.~Casoli,
     Kluwer Academic Publishers, Dordrecht, p1
 \bibitem[1999]{xyz}
   Johansson L.E.B., 1997, IAU Symp. 178, Molecules in Astrophysics,
     Probes and Processes, ed.\ E.F.~van Dishoeck, Kluwer Academic Publishers,
     Dordrecht, p515
 \bibitem[1999]{xyz}
   Johansson L.E.B., Olofsson H., Hjalmarson \AA, Gredel R., Black J.H., 1994,
     \AaA\ \vol{291}, 89
 \bibitem[1999]{xyz}
   Lada C.J., 1976, \ApJS\ \vol{32}, 603
 \bibitem[1999]{xyz}
   Larson R.B., 1981, \MNRAS\ \vol{194}, 809
 \bibitem[1999]{xyz}
   Lequeux J., Le Bourlot J., Pineau de For\^ets G., \etal, 1994,
     \AaA\ \vol{292}, 371
 \bibitem[1999]{xyz}
   Linke R.A., Goldsmith P.F., 1980, \ApJ\ \vol{235}, 437
 \bibitem[1999]{xyz}
   Mauersberger R., Henkel C., 1989, \AaA\ \vol{223}, 79
 \bibitem[1999]{xyz}
   Mauersberger R., Henkel C., Sage L.J., 1990, \AaA\ \vol{236}, 63
 \bibitem[1999]{xyz}
   MacLaren I., Richardson K.M., Wolfendale A.W., 1988, \ApJ\ \vol{333}, 821
 \bibitem[1999]{xyz}
   Millar T.J., Herbst E., 1990, MNRAS \vol{242}, 92
 \bibitem[1999]{xyz}
   Nyman L.-\AA., 1984, \AaA\ \vol{141}, 323
 \bibitem[1999]{xyz}
   Nyman L.-\AA., Millar T.J., 1989, \AaA\ \vol{222}, 205
 \bibitem[1999]{xyz}
   Nyman L.-\AA., Olofsson H., Johansson L.E.B., Booth R.S., Carlstr\"om U.,
     Wolstencroft R., 1993, \AaA\ \vol{269}, 377
 \bibitem[1999]{xyz}
   Rubio M., Garay G., Lequeux J., 1997, IAU Symp. 178 Abstract Book,
     Molecules in Astrophysics, Probes and Processes,
     eds.\ D.J.~Jansen, M.R.~Hogerheijde, E.F.~van Dishoeck, Leiden, p302
 \bibitem[1999]{xyz}
   Rubio M., Lequeux J., Boulanger F., 1993a, \AaA\ \vol{271}, 9
 \bibitem[1999]{xyz}
   Rubio M., Lequeux J., Boulanger F., \etal, 1993b, \AaA\ \vol{271}, 1
 \bibitem[1999]{xyz}
   Schilke P., Walmsley C.M., Pineau de For\^ets G., \etal. 1992,
     \AaA\ \vol{256}, 595
 \bibitem[1999]{xyz}
   Schwering P.B.W., Israel F.P., 1989, \AaAS\ \vol{79}, 79
 \bibitem[1999]{xyz}
   Sims I.R., Queffelec J.L., Defrance A., Rebrion-Rowe C., Travers D.,
   Bocherel P., Rowe B.R., Smith I.W.M., 1993, JCP \vol{100}, 4229
 \bibitem[1999]{xyz}
   Skatrud D.D., de Lucia F.C., Black G.A., Sastry K.V.L.N., 1983,
     J. Mol. Spec. \vol{99}, 35
 \bibitem[1999]{xyz}
   Turner B.E., Gammon R.H., 1975, \ApJ\ \vol{198}, 71
 \bibitem[1999]{xyz}
   van Dishoeck E.F., Black J.H., 1988, \ApJ\ \vol{334}, 771
 \bibitem[1999]{xyz}
   Watson W.D., Anicich V.G., Huntress W.T., 1976, \ApJ\ \vol{205}, L165
 \bibitem[1999]{xyz}
   Westerlund B.E., 1990, \AaAR\ \vol{2}, 29
 \bibitem[1999]{xyz}
   Wilson T.L., Rood T.R., 1994, \ARAA\ \vol{32}, 191
 \bibitem[1999]{xyz}
   Wouterloot J.G.A., Walmsley C.M., 1986, \AaA\ \vol{168}, 237
 \bibitem[1999]{xyz}
   Ziurys L.M., Snell R.L., Dickman R.L., 1989, \ApJ\ \vol{341}, 857
\end{thebibliography}
\end{document}